# The multifractal fly: a dynamically multilayered visual system

[1]M. S. Baptista, [2]Celso Grebogi and [3]Roland Köberle

[1]*Universität Potsdam, Institut für Physik Am Neuen Palais 10, D-14469, Potsdam, Deutschland;* [2]*Centre for Applied Dynamics Research, King's College, University of Aberdeen, Aberdeen AB24 3UE, UK;* [3]*Instituto de Física, Universidade de São Paulo, C. P. 369, 13560-970 São Carlos, SP, Brasil*

(Dated: August 22, 2008)

We dynamically analyze our experimental results on the motion sensitive spiking H1 neuron of the fly's visual system. We find that the fly uses an alphabet composed of a few letters to encode the information contained in the stimulus. The *alphabet dynamics* is multifractal both with and without stimulus, though the multifractality increases with the stimulus entropy. This is in sharp contrast to models generating independent spike-intervals, whose dynamics is monofractal.

The conflicting demands of variability and reliability require neurons of e.g. the sensory system to judiciously adapt their dynamics to the statistics of the stimulus[1]. If a spiking neuron has to encode *relevant* features of the stimulus, it has to fire precisely timed spikes depending on the presynaptic input generated by the instantaneous stimuli the organism receives. In this Letter, we present experimental results on the fly visual system that strongly suggest that this encoding takes place on *multilayered sets*, a characteristic of the complex nature of this system. These sets are defined in terms of symbolic sequences of letters, selected from an alphabet according to the size of spike-time intervals. Furthermore, we present strong evidence that the underlying dynamics (UD) on each layer is multifractal, suggesting thus the possibility for a chaotic-like type of encoding. Already in the non-stimulus regime, because of the multifractal dynamics, the UD is highly flexible, ready to adapt its dynamics to the statistical properties of the stimulus to be encoded. Then, in the presence of the stimulus, finely tuned spike times ride on a set whose UD has now an increased multifractality, shaped by the properties of the stimulus.

Multifractal analyses of neural activity have been performed by some authors [2, 3], using real brain data: Bershadsky et al.[2] use multifractality to analyse long term correlations in a particular region of anaesthethized rats, whereas Zheng et al.[3] envisage its use as a neurosurgical tool. Silchenko and Hu[4] study the effect of stochastic resonance in an artificial noisy bistable system. By contrast, we analyze neural sensory data by combining elements of Information Theory and Methods of the Theory of *Dynamical Systems*. We use orders of magnitude more data than could be obtained from mammalian brain studies, to be able to reveal the existence of a highly nontrivial dynamics.

We establish the underlying dynamical behavior of the neural activity in a prominent example of a spiking neuron: the H1 neuron. This neuron is located in the lobula plate of the fly Chrysomya megacephala, and is mainly sensitive to image motion associated with horizontal back-to-front rotations around a vertical axis[5]. This neuron was stimulated by a computer-controlled random, vertical bar pattern with horizontal velocity $v(t)$, new frames being shown every 2 milliseconds[6]. In order to be minimally representative, we selected four types of stimuli $v(t)$: $S_0$ = no-stimulus, $S_1$ = constant velocity, $S_2$ = a stimulus generated by an Ornstein-Uhlenbeck process with correlation time $\tau_c = 20$ milliseconds[7] and $S_3$ = an uncorrelated Gaussian stimulus.

Such a spiking neuron generates a sequence of spike times $t_i$, $i=1,2,3,\ldots N_s$, where $N_s \sim 10^6$ in our experiments. All the information received is compressed into the sequence of intervals $\Delta t_i = t_{i+1} - t_i$. In order to extract the UD, we classify all the possible intervals, depending on their size, into a discrete set of cardinality $N$: $\Delta t_i \leq d_1, d_1 < \Delta t_i \leq d_2, d_2 < \Delta t_i \leq d_3$ etc, where $d_j, j=1,2,\ldots,N-1$ are a set of *dividers*, each index $j$ generating one layer. Evidently, if we make this set large enough, we recover the original intervals. The question is: can we choose a reasonably small set of layers, without compromising the information contained in the original intervals $\Delta t_i$ ? and study their dynamics ? In other words, *what is the size of the alphabet the fly's H1 neuron uses to speak postsynaptically ?*

We choose our dividers so as to minimize information loss or maximize Shannon's entropy[8]. For a given set of $N-1$ dividers, we convert the sequence of spike intervals $\Delta t_i$, into a sequence of words of length $L$ composed of $N$ letters. Notice that an $L$-letter word may comprise a very long time interval. Now count up all the different $L$-letter words showing up in an experiment, get their probabilities $P_k$ and compute the average entropy (per letter) of a word of length $L$ with $N$ possible letters of the experimental sequence

$$H(L,N) = -\frac{1}{L}\sum_k P_k \log_2(P_k). \qquad (1)$$

| N | $S_0$ | | | | | | | $S_1$ | | | | | | | $S_2$ | | | | | | | $S_3$ | | | | | | |
|---|---|---|---|---|---|---|---|---|---|---|---|---|---|---|---|---|---|---|---|---|---|---|---|---|---|---|---|---|
| | 2 | 3 | 4 | 5 | 6 | 7 | 8 | 2 | 3 | 4 | 5 | 6 | 7 | 8 | 2 | 3 | 4 | 5 | 6 | 7 | 8 | 2 | 3 | 4 | 5 | 6 | 7 | 8 |
| $d_1$ | 23 | 9 | 9 | 9 | 9 | 9 | 6 | 6 | 6 | 4 | 4 | 4 | 4 | 3 | 5 | 3 | 3 | 3 | 3 | 3 | 2 | 5 | 5 | 4 | 3 | 3 | 3 | 3 |
| $d_2$ | - | 23 | 23 | 23 | 23 | 14 | 9 | - | 10 | 6 | 6 | 6 | 5 | 4 | - | 5 | 5 | 4 | 4 | 4 | 3 | - | 9 | 5 | 4 | 4 | 4 | 4 |
| $d_3$ | - | - | 49 | 49 | 34 | 23 | 14 | - | - | 10 | 10 | 8 | 6 | 5 | - | - | 11 | 5 | 5 | 5 | 4 | - | - | 9 | 5 | 5 | 5 | 5 |
| $d_4$ | - | - | - | 71 | 49 | 34 | 23 | - | - | - | 15 | 10 | 8 | 6 | - | - | - | 11 | 11 | 7 | 5 | - | - | - | 9 | 9 | 6 | 6 |
| $d_5$ | - | - | - | - | 71 | 49 | 34 | - | - | - | - | 15 | 10 | 8 | - | - | - | - | 34 | 11 | 7 | - | - | - | - | 21 | 9 | 7 |
| $d_6$ | - | - | - | - | - | 71 | 49 | - | - | - | - | - | 15 | 10 | - | - | - | - | - | 34 | 11 | - | - | - | - | - | 21 | 9 |
| $d_7$ | - | - | - | - | - | - | 71 | - | - | - | - | - | - | 15 | - | - | - | - | - | - | 34 | - | - | - | - | - | - | 21 |

TABLE I: The best time dividers (rows 3,4, ...) in units of bins, which generate the entropies $H_2$ for all data sets, considering different number of letters $N$ ($2^{nd}$ row). Since $H(L=1, N) \geq H(L>1, N)$, we calculate these dividers fixing $L=1$.

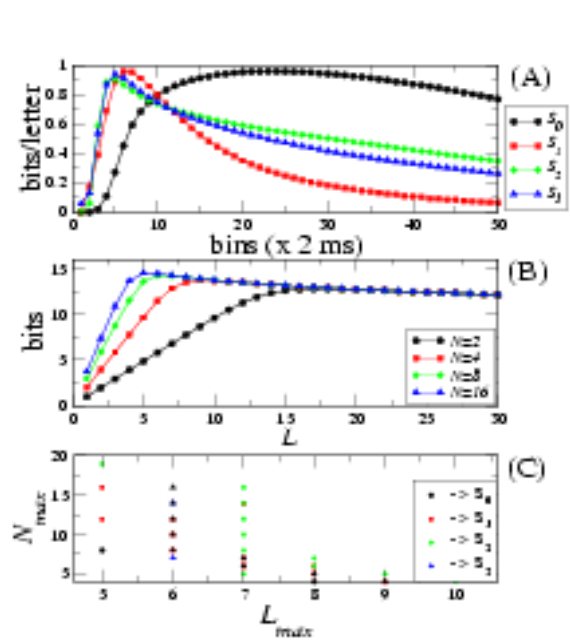

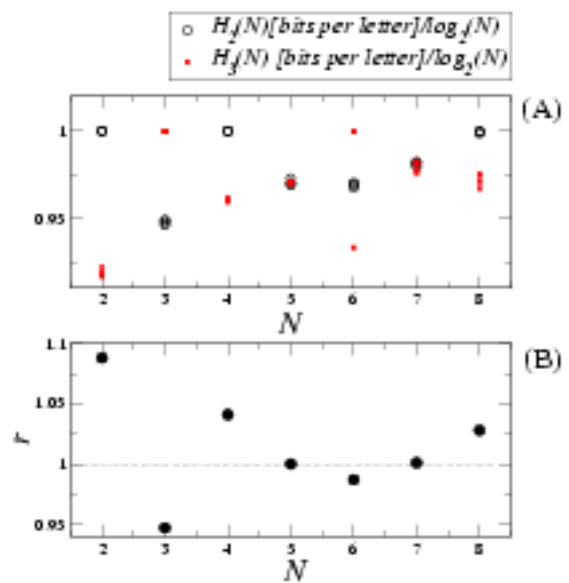

FIG. 1: (Color online) (A) Entropy vs. divider for a 2-letter code, for the four data sets: $S_0, S_1, S_2, S_3$. The entropy maximum gives the best divider $d_1$. (B) Word-Entropy $(H_{max}(L, N) \times L)$ vs. the length of words $L$ for the data $S_1$. Note, that for each $N$ there is a value of $L$ for which the entropy reaches its maximum. We refer to these N and L values as $N_{max}$ and $L_{max}$. Each curve corresponds to a different number of letters. (C) Scaling law that relates $N_{max}$ with $L_{max}$ for which the maximum of the entropy is achieved, for all the data sets.

FIG. 2: (Color online) (A) Entropy per letter normalized by $log_2(N)$ for all data sets. Circles represent the entropy $H_2(N)/log_2(N)$ and squares the entropy $H_3(N)/log_2(N)$ for all data sets. (B) Ratio of $r = <H_2>/<H_3>$, where $<>$ indicates the average over $S_0, ..., S_3$.

Figure 1 (A) shows the entropy for $L = 10$ with only one divider, for the four different data sets. The entropy shows a maximum at $[d_1(S_0), d_1(S_1), d_1(S_2), d_1(S_3)] = [23, 6, 5, 5]$ bins. We now construct a uniquely defined generating[10] partition of our time intervals, i.e., our alphabet has to be able to allow a 1-to-1 encoding of sequences of intervals. This will yield the partition with $N = 2 \times 2^J = 2, 4, 8, \ldots$. In order to be generating all the dividers of a particular layer $J$ must already be contained in the previous layers $j = 1, 2, \ldots, J-1$. With this consistency requirement in mind, we construct Table I in two steps. Firstly, maximizing the entropy $H_2(N = 2 \times 2^J)$ for $N = 2, 4, 8, \ldots$, we get the respective dividers in Table I. Secondly, we turn to the remaining values of $N$, searching for dividers satisfying our consistency condition. This is indeed possible, if we fix, e.g., the bold faced dividers and perform a constrained maximization to obtain the remaining ones. We obtain the entropy per letter $H_2(N = 3, 5, 6, 7, ..)$, which equals to within 5 % the entropy obtained with an unconstrained maximization, which yields completely different dividers. We conclude that, $H_2(N)$ for $N = 2 \times 2^J = 2, 4, 8, \ldots$ is a generating partition. Is it unique ? If we perform the same procedure for other partitions, e.g., $N = 3 \times 2^J = 3, 6, 12, \ldots$, the first step yields the same value for the unconstrained entropy maxima $H_3(N = 3, 6, 12, \ldots)$. In the second step, to obtain the remaining dividers, the constrained maximizations yield smaller entropies $H_3(N = 2, 4, 5, 7, 8, ..)$ by as much as 10 % as compared to the unconstrained ones, as we show in Fig.2. Here,

$H_3(N = 2)$ means: (i) - maximize $H_3(N = 3)$ to get $[d_0, d_1]$; (ii) - maximize $H_3(N = 2)$, using either $d_0$ or $d_1$, whichever gives the larger entropy. The only exception here is one point corresponding to $N = 3$. The same is also true for other partitions, starting with prime numbers 5, 7, indicating thus, that within errors the binary partition is the only generating one.

Figure 1 (C) shows that the number of letters needed to maximize the entropy, decreases with increasing word-length $L$, due to correlations manifesting themselves and undersampling effects. Since these put a limit to the size of the alphabet usable, we address the UD using just a few letters. We notice, however, that even for $N$ smaller than $N_{max}$, e.g seven dividers for experiments $S_2$ and $S_3$, we obtain essentially the information the original spike trains convey about the stimulus[8]. Fig.3 shows the information per letter $I_A$ for $S_2$[11], divided by the information the original spike trains conveys about the stimulus, using the dividers of Table I. I.e., for $N = 2, 3, 4, 5 \ldots$, we use the dividers $[5], [3, 5], [3, 5, 11], [3, 4, 5, 11], \ldots$

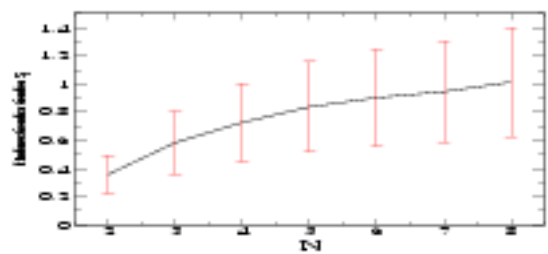

FIG. 3: (Color online) Information per letter $I_A$ the alphabet conveys about the stimulus for $S_2$, divided by the average Information per spike $I_S$: $f = I_A/I_S$. The computation of the noise-entropy $H_{Noise}$, where $I_A = H - H_{Noise}$, introduces the large error bars. Our data are extrapolations to $L \to \infty$[9].

Since we are going to extrapolate below to $L$ large, we focus on sequences of words of length $L$ containing four letters, using the best dividers of Table I. Only fine details of our results depend on this choice.

We now encode each of these L-letter words in a one-to-one map into a real number $W_i$, $0 \le W_i \le 1$, whose frequencies we count in order to compute their probabilities $p_i$. The structure of the space of sequences $\mathcal{C}(W)$ for a given $N$ may now be uncovered by computing the generalized dimensions $D_q$[12, 13]. These are logarithmic ratios between the probabilities $p_i$ and their physical occupation $e$, which in our space $\mathcal{C}$, is give by $e = N^{-L}$. The index $q$ can be thought of as a filter: a larger $q$ enhances this ratio for large probabilities, whereas a negative one emphasizes the smaller probabilities:

$$D_q = \lim_{L \to \infty} \frac{\ln(\sum_i p_i^q)}{(1-q) L \ln(N)} \quad (2)$$

An equivalent quantity is $f_\alpha$, the Legendre transform of $(1-q)D_q$. The index $\alpha$ measures the possible local fractal dimensions[14] in our space $\mathcal{C}$, occurring with singularity strength $f_\alpha$. This is the global dimension of the set of points, that locally scales with singularity strength $\alpha$. They are given by $\alpha = \frac{d[(q-1)D_q]}{dq}$, $f_\alpha = q\alpha - (q-1)D_q$[15].

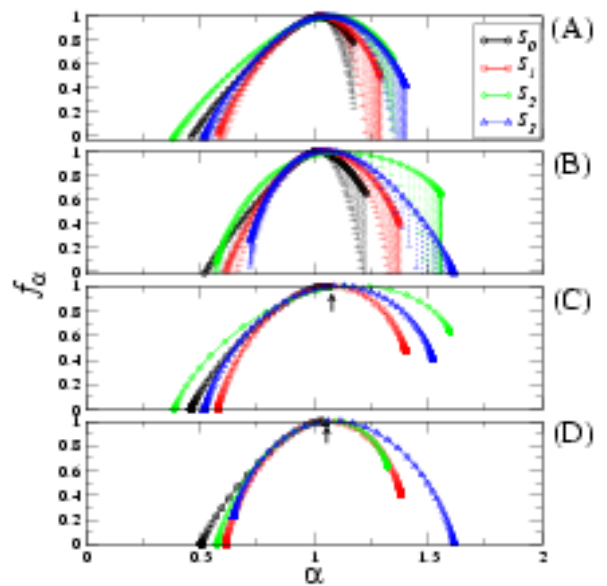

FIG. 4: (Color online) $f_\alpha$ spectra using Table I: (A) two letters and (B) four letters. To avoid cluttering, we show only one-sided error bars as dashed lines. Spectra analogous to (A) and (B), using Eq.(3): (C) for 2 letters with $b_{S_0:S_1:S_2:S_3} = [30; 1.55; 2.0; 1.44]$ and (D) for 4 letters with $b_{S_0:S_1:S_2:S_3} = [30; 1.45; 2.0; 1.2]$. The arrows indicate the endpoint $\alpha_{max}$ of the curve for $S_0$.

In Fig.4(A-B) we show the spectra of the symbolic sequences in $\mathcal{C}(W)$ with their error[15], which allow us to draw the following conclusions. The H1 neuron has a multifractal character, exhibiting the existence of an infinite number of dimensions $\alpha$ with densities $f_\alpha$. This is in sharp contrast to a memoryless, uncorrelated spike train with a Poissonian or similar probability distribution. In fact any distribution with independent increments yields a nontrivial $f_\alpha$ for suboptimal dividers, but it is really monofractal for the optimal ones with $D_q = log_2(N)$. In fact for $N = 2$, let $p_0/p_1$ be the probabilities for an interval being smaller/greater than $d_1$. For a Poisson distribution, the probability for no spike in $[0, t]$ - or equivalently all spike-intervals larger than $t$ - is $e^{-\lambda t}$, where $\lambda$ is the spike-rate. Therefore $p_1(Poisson) = e^{-\lambda d_1}$. Then $D_q = log_2(p_0^q + p_1^q)/(1 - q)$. The maximization of the entropy obviously gives the uniform distribution $p_0 = p_1 = 1/2$ and $e^{-d_1(Poisson)\lambda} = 1/2$, yielding $D_q = log_2(2)$. Real spike trains by contrast are multifractal even for the best dividers for all types of stimuli. The fractal dimensions are roughly the same for all the data sets, since $D_0 = f_1 = 1$. This means that the neuron's dynamics has continuous support on $\mathcal{C}$, the probability measure being distributed without "holes". The spectra's shape - $f_{\alpha_{min}} = 0$ and $f_{\alpha_{max}} > 0$ (except for $S_3$)

- indicates, that $\mathcal{C}$ has at least two scales and two main components: high probability sets - hot spots, localized in small portions of the the symbolic space $\mathcal{C}$ with density $f_{\alpha_{min}}$, and another low probability set - cold spots, spread out all over $\mathcal{C}$ with density $f_{\alpha_{max}}$. In the set associated with $S_0$, the number of cold spots very much dominates - $f_{\alpha_{max}} \sim \alpha_{max} \sim 1$ - the hot ones, implying one dominant scale. For the other data sets, the scales are comparable. Therefore, as the dynamics of the stimulus becomes faster as we go from $\tau_c = \infty$ to $\tau_c = 0$, the suppressed scale emerges.

Given the $f_\alpha$ spectra of our data, what is the simplest set with this spectrum ? To address this question, we construct a probabilistic 2-scale set with the following rule: an interval of length unity is divided into $b+1$ equal pieces, such that one piece receives $p_0$ of the original measure and the remaining $b$ receive $bp_1$, with $p_0 + bp_1 = 1$. Iterating this process self-similarly yields a set of dimension

$$D_q = \frac{ln(p_0^q + bp_1^q)}{(1-q)ln(1+b)}, \quad (3)$$

with $p_0 = (1+b)^{-\alpha_{min}}$ and $p_1 = (1-p_0)/b$, where $b$ is adjusted to produce the correct value for $f_{\alpha_{max}}$ in Fig.4(A-B). The resulting $f_\alpha$ spectra are shown in Fig.4(C-D). In (D), to reproduce the spectra for $S_3$ in (B), we make $p_0 = (1+b)^{-\alpha_{max}}$.

Notice, that $b$ jumps to lower values, once a stimulus is turned on. This means, that for no-stimulus the dynamics distributes the measure rather uniformly into 31 equal intervals, where 30 of them receive $p_1$ of the measure and one receives $p_0$ with $p_0 > p_1$. At the $r^{th}$ iteration of the set, 30 intervals will contain $p_1^r$ of the measure (cold spots), one interval will contain $p_0^r$ of the measure (hot spots), i. e., virtually all of the measure, and the rest will receive combinations proportional to $p_0^{r-k} p_1^k, 0 \leq k \leq r$. When stimulated, the number of pieces drops dramatically to $\sim 1$, the hot spots approximately balancing the cold ones. The stimuli thus reshape the probability landscape, which becomes more and more structured as the stimulus entropy increases.

In conclusion, the *underlying dynamics* for the H1 neuron can be extracted by a finite sized alphabet with about four letters. Analyzing sequences written in this alphabet, allows us to exhibit the multifractal character of the sequence-space. Here the stimulus shapes the landscape from mainly uniform to highly structured as the stimuli become increasingly dynamically variable. But this reshaping is played out on different layers, which are dynamically linked - a behaviour typical of complex systems. The tools developed here for the analysis of our data are of general applicability. They reveal a fascinating complexity in the dynamics of the fly's visual system, even in the basic aspects of the spike generating dynamics, and they provide a means for their manipulation and control. It remains to be seen, whether the layered structure uncovered here, has a counterpart in the fly's optical information processing system.

Acknowledgements: Thanks go to the FAPESP (N. S. B., R. K., C. G.) and the A. v. Humboldt Foundation (N. S. B.) for partial support. It is R. K.'s pleasure to thank W. Bialek and R. R. van Steveninck for their infinite patience, when sharing their insights.